\documentclass[USenglish,11pt]{article}

\usepackage{hyperref}   
\usepackage{subfigure}  

\usepackage{newpxtext, newpxmath}

\usepackage[utf8]{inputenc}
\usepackage{todonotes}
\usepackage{qcircuit}
\usepackage{graphicx}

\newcommand{\ket}[1]{\ensuremath{|#1\rangle}}
\newcommand{\bra}[1]{\ensuremath{\langle#1|}}
\newcommand{\ketbra}[2]{\ensuremath{\ket{#1}\bra{#2}}}
\newcommand{\braket}[2]{\ensuremath{\langle{#1}|{#2}\rangle}}
\newcommand{\proj}[1]{\ensuremath{\ket{#1}\bra{#1}}}

\newcommand{\C}{\mathbb{C}}

\newcommand{\myCirc}[1]{\mbox{\Qcircuit @C=1.em @R=1.2em {#1}\null\mbox{}} }

\title{Quantum distance-based classifier with constant size memory, distributed
	knowledge and state recycling}
\author{Przemys{\l}aw Sadowski\footnote{psadowski@iitis.pl}}
\date{Institute of Theoretical and Applied Informatics,\\Polish Academy of
	Sciences,\\Gliwice, Poland\\{\vspace{.5cm}}
	2 March 2018}

\begin{document}
\def\test{\ket{\psi_{\mathrm{test}}}}
\def\psim{\ket{\psi_i}}

\maketitle
\begin{abstract}
In this work we examine recently proposed distance-based classification method
designed for near-term quantum processing units with limited resources. We
further study possibilities to reduce the quantum resources without any
efficiency decrease. We show that only a part of the information undergoes
coherent evolution and this fact allows us to introduce an algorithm with
significantly reduced quantum memory size. Additionally, considering only
partial information at a time, we propose a classification protocol with
information distributed among a number of agents. Finally, we show that
the information evolution during a measurement can lead to a better solution
and that accuracy of the algorithm can be improved by harnessing the state after
the final measurement.
\end{abstract}

\section{Introduction}

Recently, a significant effort has been made to develop machine learning
techniques that harness the power of quantum processing units.
It is foreseen that quantum mechanics will offer tantalizing prospects to
enhance machine learning, ranging from reduced computational complexity to
improved generalization performance~\cite{biamonte2017quantum, schuld2015introduction}.
A number of methods have been proposed, with 
algorithms for cluster finding~\cite{lloyd2013quantum}, principal component analysis, quantum support vector machines, and
quantum Boltzmann machines being the most promising ones~\cite{biamonte2017quantum}.

The progress in the field is very dynamic, but the devices available in the
near-term provide limited resources to harness the proposed methods.
Thus, a question of critical importance for  the  field  of  quantum  computing,
that will only gain more interest in the near future is
whether quantum processing units can provide resources for algorithms solving
well-defined computational tasks that lie beyond the reach of state-of-the-art
classical computers.

A number of approaches to harness near-term devices have appeared.
One of the most important directions of research is studying applications of
sampling tasks, with works stating that some sampling tasks must take exponential time in a classical
	computer~\cite{boixo2016characterizing}.
Another promising field that could yield valuable, noise-robust methods 
is concerned with quantum approximate optimization algorithms, designed to run on a gate
	model quantum computer and has shallow depth~\cite{farhi2016quantum}.
Also, quantum annealing is argued to be reliable enough for problems unfeasible
for classical computers~\cite{smelyanskiy2012near}.
An important trend in context of this work is trying to get to the regime that 
existing  supercomputers  cannot 
easily  simulate with methods to scale up using only small trusted
devices~\cite{wiebe2014hamiltonian}.
This results encourage further study of methods that focus on near-term resources
more that sole scaling properties.

In the field of near-term quantum machine learning some results demonstrate that, while complex fault-tolerant
architectures will be required for universal quantum computing, a significant
quantum advantage already emerges in existing noisy
systems~\cite{riste2017demonstration}.
The approach that is of particular importance in context of this work,
is recently proposed distance-based classification method designed for low depth
quantum interference circuits~\cite{schuld2017implementing}.

The main goal of this work is to consider ways to improve distance-based
classification in terms of needed resources and to develop new features that
extend its usefulness as well as boost its efficiency.

\subsection{Quantum distance-based classifier}
A distance-based classifier we consider in this paper is an example of a kernel
method~\cite{elisseeff2002kernel}.
In this method classification, or prediction of the class label of some given
test sample is determined based on a value of similarity function, called a
kernel.
The similarity is measured between the test sample and all of the training
samples. For instance, having a collection of training samples
$\{(x_i,y_i)\}_{i=1,\ldots,M}$
consisting of pairs of feature vectors $x$ and corresponding class labels
$y_i\in\{-1,1\}$ the goal is to assign a class $y'$ to a test sample $x'$. The
classifier we consider may be seen as a kernelized binary classifier 
\begin{equation}
y'(x') = \mathrm{sgn}\sum_{i=1}^M y_i\kappa(x_i, x'),
\end{equation}
where the similarity is measured based on the distance
\begin{equation}
\kappa(x, x') = 1-\frac{1}{4M}|x-x'|^2,
\end{equation}
with $M$ being a normalizing constant equal to the size of the training set.

The quantum circuit implementation of the classifier is based on encoding the
feature vectors in quantum states. For normalized feature vectors $x_i$ the
corresponding quantum states are 
\begin{equation}
\ket{\psi_i} = \sum x_{i,j}\ket{j}.
\end{equation}
The classification is performed by preparing a state that encodes both the
training set and the test sample and measuring the state after a simple
pre-processing.
For this purpose we entangle an ancillary qubit with the data.
Additionally we introduce registers for element index $i$ and its class $y_i$
and obtain a state
\begin{equation}
A\sum_i\ket{i}\ket{\psi(i)}\ket{y_i},\label{eq:the-state}
\end{equation}
where
\begin{equation}
\ket{\psi(i)} = \frac{1}{\sqrt{2}}(\ket{0}\test+\ket{1}\ket{\psi_i})
\end{equation}
and $A = \frac{1}{\sqrt{M}}$.
The only quantum operation we perform is a Hadamard operation on the ancillary
register.
Finally we obtain a state
\begin{equation}
A\sum_i\ket{i}\otimes\left((H\otimes1)(\ket{\psi(i)})\right)\otimes\ket{y_i}
,\label{eq:pre-measurement}
\end{equation}
where
\begin{equation}
(H\otimes1)(\ket{\psi(i)}) =\frac{1}{\sqrt{2}}
\left(
\ket{0}\ket{\psi^{+}_i}
+
\ket{1}\ket{\psi^{-}_i}
\right),
\end{equation}
and $\ket{\psi^{\pm}_i}=\test \pm \ket{\psi_i}$.
If we perform a measurement on the resulting state and get $\ket{0}$ as the result on
the auxiliary register, the conditional probabilities for each
of the classes $y$ are proportional to $\sum_{y_i=y}|x_i+x_{\mathrm{test}}|^2$.
For any further details we encourage the reader to see~\cite{schuld2017implementing}.

\section{Distance-based classifier as a quantum channel}
In this section we aim to show that the resources used to implement the distance-based
classifier as described in the previous section can be significantly reduced.
The quantum operations we do during classification apply only to the ancillary
and feature vectors registers. This means that these are the only registers that
need to be encoded in the quantum memory. The index and the class registers may
be classical. Thus, the algorithm may be expressed as a quantum channel,
significantly reducing the number of required qubits.

\subsection{Mapping the classifier to a quantum channel}
Using a density matrix description that allows us to mix quantum and classical
registers 
we can write a state encoding the training set similar to the one in equation
(\ref{eq:the-state}).
The state can be represented as
\begin{equation}
\frac{1}{M}\sum_i \ketbra{i}{i}\otimes\proj{\psi(i)}\otimes\ketbra{y_i}{y_i},
\end{equation}
where $\ket{\psi(i)} = \frac{1}{\sqrt{2}}(\ket{0}\test+\ket{1}\ket{\psi_i})$.
Thus, in order to perform classification it is sufficient to prepare a state
$\rho_i = \proj{\psi(i)}$ for $i$ chosen from uniform distribution
and apply Hadamard operation on the auxiliary register
\begin{equation}
\Phi(\rho_i) = (H\otimes 1) \rho_i (H \otimes 1)^\dagger.\label{eq:mixed-pre-measurement}
\end{equation}
If the result is 0, the output class is $y_i$, the same as for the training
element.
As the auxiliary and feature registers are the only ones entangled, the
probability distribution is exactly the same.

Note that the channel description is useful for representation of the whole
classification procedure. The key part that is done on a QPU can still be
described as a unitary operation on a reduced state.
Thus, we can design the whole procedure to be as follows.
\begin{enumerate}
\item For given test sample compute normalized vector $x_{\mathrm{test}}$.
\item Pick random training sample $i$ from a uniform distribution.
\item Prepare a circuit that prepares the comparison state $(H\otimes1)\ket{\psi(i)}$.
\item Run the QPU, preparing $(H\otimes1)\ket{\psi(i)}$ and measure the resulting state.
\item If the result at the auxiliary register is $\ket{1}$, goto 2.
\item Return $y_i$.
\end{enumerate}
The mixture of classical and quantum steps is corresponding to the quantum
channel description.

\subsection{Probability distribution equivalence}
To show the equivalence of the probability distribution we recall the
probability of measuring
the state in Eq.~(\ref{eq:pre-measurement}) and obtaining result 0 on the
auxiliary register and $y$ on the class register
\begin{equation}
p(0, y)= \sum_{l,j} \left|(\bra{l}\bra{0}\bra{j}\bra{y})\left(A\sum_i \ket{i}\otimes(\ket{0}\ket{\psi^+_i}+\ket{1}\ket{\psi^-_i})\otimes\ket{y_i}\right)\right|^2=
\end{equation}
\begin{equation}
\frac{1}{M}\sum_{y_i=y,j} \left|\braket{j}{\psi^+_i}\right|^2.
\label{eq:basic-scheme-measurement-probability}
\end{equation}
It is straightforward to check that a measurement given state in Eq.~(\ref{eq:mixed-pre-measurement}) results in the same probability distribution.
For a single sampled $i$ one obtains the probability of successful post-selection equal to
\begin{equation}
\mathrm{Tr}\left( (H\otimes 1) \proj{\psi(i)} (H \otimes 1)^\dagger (\ketbra{0}{0}\otimes 1) \right)=
\end{equation}
\begin{equation}
\mathrm{Tr}\left( (\ket{0}\ket{\psi^+_i}+\ket{1}\ket{\psi^-_i})(\bra{0}\bra{\psi^+_i}+\bra{1}\bra{\psi^-_i}) (\ketbra{0}{0}\otimes 1) \right)=
\end{equation}
\begin{equation}
\mathrm{Tr}\left( \proj{\psi^+_i} \right)=\sum_j |\braket{j}{\psi^+_i}|^2.
\end{equation}
If we consider the probability of sampling $i$ equal to $\frac{1}{M}$ and sum over
all of the samples with given class label $y_i$ equal to $y$ the result matches Eq.~(\ref{eq:basic-scheme-measurement-probability}).

\subsection{Implementation}

For each comparison of randomly picked sample label $m$ with the test sample, we
consider sample state $\psim$ and the input state $\test$. The main challenge in
implementing the algorithm on a QPU is to design proper quantum
circuit that prepares the desired state. In this case, we aim at preparing
$\frac{1}{\sqrt{2}}(H\otimes 1)(\ket{0}\otimes\test+\ket{1}\otimes\psim)$. 

We begin with simplified 2 features case. 	
Let us fix $\test = \cos(\varphi)\ket{0} + \sin(\varphi)\ket{1}$ and
$\ket{\psi_{\mathrm{m}}} = \cos(\phi_m)\ket{0} + \sin(\phi_m)\ket{1}$.
We present a general formula for preparing the state for any test and training
samples. For preparing the desired state it is necessary to prepare a state
with relative phase dependent on the auxiliary register. We assume that the
initial state is $\ket{0}\otimes\ket{0}$. The initial rotation is performed
along $y$ axis by angle equal to $\alpha_1 = \pi/4+(\varphi-\phi_m)/2$. Then
controlled negation is performed to obtain different relative phases depending
on the value of the auxiliary register. The final rotation, again along $y$
axis by $\alpha_2 = -\pi/4+(\varphi+\phi_m)/2$ results in the desired state.
The complete preparation operator is equal to
\begin{equation}
(H\otimes R_y(\alpha_2)) CNOT (H\otimes
R_y(\alpha_1)) (\ket{0}\otimes\ket{0}).
\end{equation}
The resulting circuit can be expressed as

\begin{center}
	\mbox{\myCirc {
			A& & \gate{H}     & \ctrl{1}     & \gate{H}   & \measure{\mathbf{M}} \\
			F& & \gate{R_y(\alpha_1)} & \gate{\oplus}& \gate{R_y(\alpha_2)} &\qw }}
\end{center} \vspace{1ex}
where \textbf{M} stands for final projective measurement.

The method generalizes to larger feature registers provided that all the
necessary controlled operators are provided in the architecture. If there are
some restrictions it is important to note that in case of larger register, the
states are entangled and preparing complexity grows, because we cannot prepare
each of the qubits separately. In such case a good candidate for searching for
proper preparation circuits is to develop a heuristic method for finding the
circuits that with some smart guess of the available gate set finds the circuit
without the need to provide explicit algorithm~\cite{sridharan2010reduced,sadowski2013geneticIJQI}.

The circuit that prepares desired state for any training and test samples with 4 features
consists of four rotations and three controlled negations. The circuit is in the
following form
\begin{center}
	\mbox{\myCirc {
			A& & \gate{H}      & \ctrl{2}    & \qw          & \qw       &\qw         
			&\ctrl{2}  &\gate{H}     &\measure{\mathbf{M}} \\
			F& & \gate{H}      & \qw         & \qw          & \ctrl{1}  &\qw         
			&\qw       &\qw          &\qw\\
			F& & \gate{R_y(1)} & \qw\oplus   & \gate{R_y(2)}& \qw\oplus
			&\gate{R_y(3)}&\qw\oplus &\gate{R_y(4)}&\qw }}
\end{center} \vspace{1ex}
where the features register consists of the two lowest qubits.

\subsection{Example}
In order to show the increased potential of the quantum distance-based
classifier we perform experiment on the same data set as
in~\cite{elisseeff2002kernel}, but using all of the features the data poses.
The introduced method additionally allows one to use arbitrarily many training
samples. Thus, we perform leave one out cross validation using the whole dataset.

The probability that post-selection is successful for given test/training class
is presented in Table~\ref{fig:results-postselection}. We observe that even with
increased number of classes considered the probability of the correct class is
always the highest. The overall resulting probability distribution of output class labels, after post-selection,
for given training sample is presented in
Table~\ref{fig:results-cross-validation}. In every case the probability of the
correct class is the highest. Overall success probabilities are given in
Table~\ref{fig:success-corss-validation}.

\begin{table}[h]
	\begin{center}
		\begin{tabular}{|c|c|c|c|}
			\hline
			\multicolumn{4}{|c|}{4 features} \\ \hline
			Class & A & B & C  \\ \hline
			A & 0.97 & 0.38 & 0.08 \\ \hline
			B & 0.38 & 0.73 & 0.66\\ \hline
			C & 0.08 & 0.66 & 0.89\\ \hline
		\end{tabular}
	\end{center}
	\caption{Average successful post-selection probability from leave one out cross-validation on the Iris dataset}\label{fig:results-postselection}
\end{table}

\begin{table}[h]
	\begin{center}
		\begin{tabular}{|c|c|c|c|}
			\hline
			\multicolumn{4}{|c|}{4 features} \\ \hline
			Class & A & B & C  \\ \hline
			A & 0.68 & 0.27 & 0.06 \\ \hline
			B & 0.22 & 0.41 & 0.37\\ \hline
			C & 0.05 & 0.40 & 0.55\\ \hline
		\end{tabular}
	\end{center}
	\caption{Output class probability distribution, conditional after
	post-selection from leave one out cross-validation on the Iris
	dataset}\label{fig:results-cross-validation}
\end{table}
\begin{table}[h]
	\begin{center}
		\begin{tabular}{|c|c|c|c|}
			\hline
			\multicolumn{4}{|c|}{4 features} \\ \hline
			Class & A & B & C  \\ \hline
			All & 0.68 & 0.41 & 0.55 \\ \hline
		\end{tabular}
	\end{center}
	\caption{Average success probability from leave one out cross-validation on the Iris dataset}\label{fig:success-corss-validation}
\end{table}

\section{Generalized measurements sequence classifier}
In this section we move to discussing further possibilities that come from the
fact that part of the system may be seen as a collection of classical registers.
In particular we aim at modeling the classification protocol as a sequence of
steps in time that results in the measurement properties consistent with
training data. Using the channel description we expressed the algorithm as a single
preparation and measurement routine.
If we want to describe a sequence of such steps where the state may be preserved
after the measurement and the action is determined
by the result of the measurement, a very convenient model is the Open Quantum
Walk (OQW)~\cite{attal2012open}, modeling any quantum system homogeneous in time
with one classical property, called position, that governs the dynamics rules.
This model enables one to clearly distinguish classical and quantum information
and describe the dynamic rules.

\subsection{Classification with distributed knowledge}
The simplest open quantum walk on a graph of size $n$ and internal states of
size $k$ is defined as a quantum channel $\Phi\in L(L(\C^{\otimes nk}))$, such that the state of
the system evolves by applying the walk channel

\begin{equation}
\rho_{t+1} = \Phi(\rho_t),
\end{equation}
where the walk channel using Kraus representation is a sum of transition
operators
\begin{equation}
\Phi(\rho) = \sum_{i,j=1}^n (\ketbra{j}{i}\otimes K_{i,j})\rho
(\ketbra{j}{i}\otimes K_{i,j})^\dagger,
\end{equation}
with $\sum_j K_{i,j}^\dagger K_{i,j}=1$. Note that $K_{i,j}\ne0$ are responsible
for
connectivity.

Because the index register could be classical it can be a position in an open
quantum walk. A single node would correspond to a single test, comparing the
test state with one fixed training element. To ensure proper information
distribution. One may see this as a training done by walking.
The class of the position measured would be the output.

The walk channel corresponding to the classifier that would implement
preparation of the training state $\ket{\psi(j)}$ during transition from
position $i$ to  position $j$ is
\begin{equation}
K_{i,j} = \frac{1}{\sqrt{d_{\mathrm{out}}(i)}} \sum_{d=1}^k \ketbra{\psi(j)}{d},
\end{equation}
where $d_{\mathrm{out}}(i)$ is number of outgoing edges from node $l$.
Alternatively, the transition operators can be described without including the
test state into the Kraus operators
\begin{equation}
K'_{i,j} = \frac{1}{\sqrt{d_{\mathrm{out}}(l)}} \sum_{f=1}^{k/2}(
\ketbra{0}{0}\otimes 1 + \ketbra{1}{1}\otimes\ketbra{\psi_j}{f}),\label{eq:walk-kraus}
\end{equation}
resulting in a scenario where the test state is introduced only once as a
quantum state, making it impossible to save it for future use without disturbing
the protocol.

As a result we obtain a procedure with distributed knowledge about the training
samples. The scenario may include many parties, where each party  has limited
knowledge that is not shared and the result is the same as in the case when
all the knowledge was available.
The procedure that implements this walk in a situation with many parties could be
as follows.
\begin{enumerate}
	\item For given test sample compute normalized vector $x_{\mathrm{test}}$.
	\item Provide one of the agents $i$ with a state $\frac{1}{\sqrt{2}}(\ket{0}\otimes\ket{\psi_{\mathrm{test}}}+\ket{1}\otimes\ket{1}\})$.
	\item The first agent $i$ applies $K'_{i,i}$ as in Eq.~(\ref{eq:walk-kraus}).
	\item Until the stop condition is reached do:
	\begin{itemize}
		\item Pick random neighbor $j$ from a uniform distribution.
		\item Transfer the state $\ket{\psi}$ applying $K'_{i,j}$ in the process.
	\end{itemize}
	\item Measure the final state $(H\otimes1)\ket{\psi}$.
	\item If the result at the auxiliary register is $\ket{1}$, goto 2.
	\item Return $y_j$.
\end{enumerate}

The caveat is that some normalization
information has to be shared to ensure proper state preparation. As Kraus
operators
change the internal state, for many training states at one site
we would have to use generalized OQW~\cite{pawela15generalized}. The model
allows to study
limiting properties of classification results~\cite{konno2013limit,sadowski2014central, attal2015central}. The
dynamics need to be
fair in some sense that is different than in the classical
case~\cite{miszczak2014magnus}.

\subsection{Distributed knowledge example}
In this case the main aim of the simulation is to show what is the influence of
the walk graph on the classification results. In the case when the graph is
complete or when the initial position is chosen randomly we would obtain exactly
the same results as for a simple channel model.

In our experiment we compare two cases chosen as representative extreme cases.
We choose our graph to be a cycle, but we consider two arrangements of the nodes.
\begin{enumerate}
	\item 150 nodes corresponding to training samples from 3 classes (A, B, C)
arranged in three clusters (AAA...BBB...CCC...), starting position in the
middle of one of the classes.

	\item 150 nodes corresponding to training samples from 3 classes (A, B, C)
	arranged in a regular pattern with neighbors being always from different
classes (ABCABC ... ), starting position in the middle of one of the cycle.
\end{enumerate}
For each of the scenarios we study the probability of correct classification.
Regardless of nodes distribution the walk will always converge to the uniform
distribution~\cite{BEDNARSKA200321}, thus exhibiting exactly the same probabilities as the channel
model. The subject of the study is the behavior during the convergence time.

\begin{figure}%
	\subfigure[Scenario 1 -- united configuration]{\includegraphics[width=0.5\textwidth]{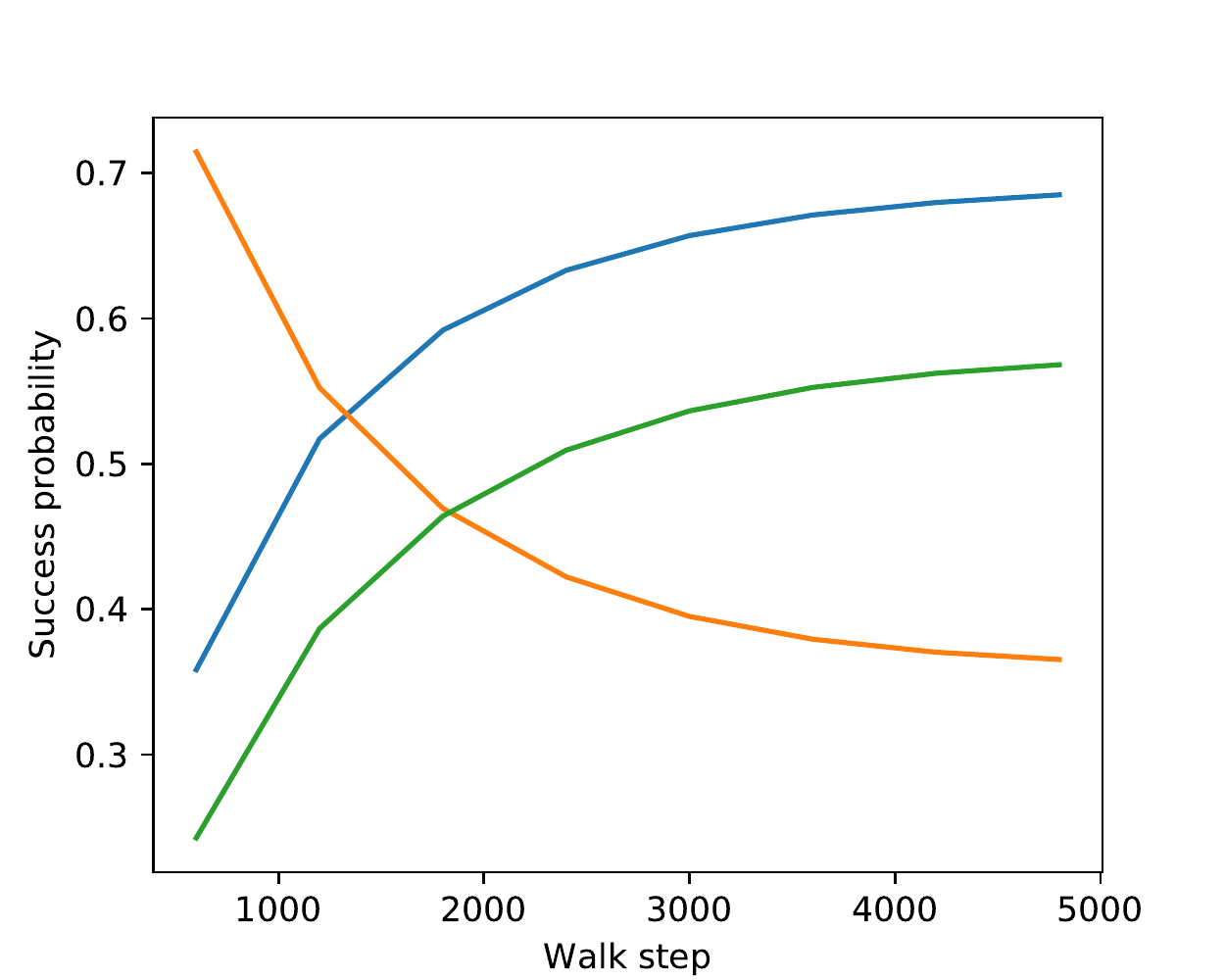}}~%
	\subfigure[Scenario 2 -- mixed configuration]{\includegraphics[width=0.5\textwidth]{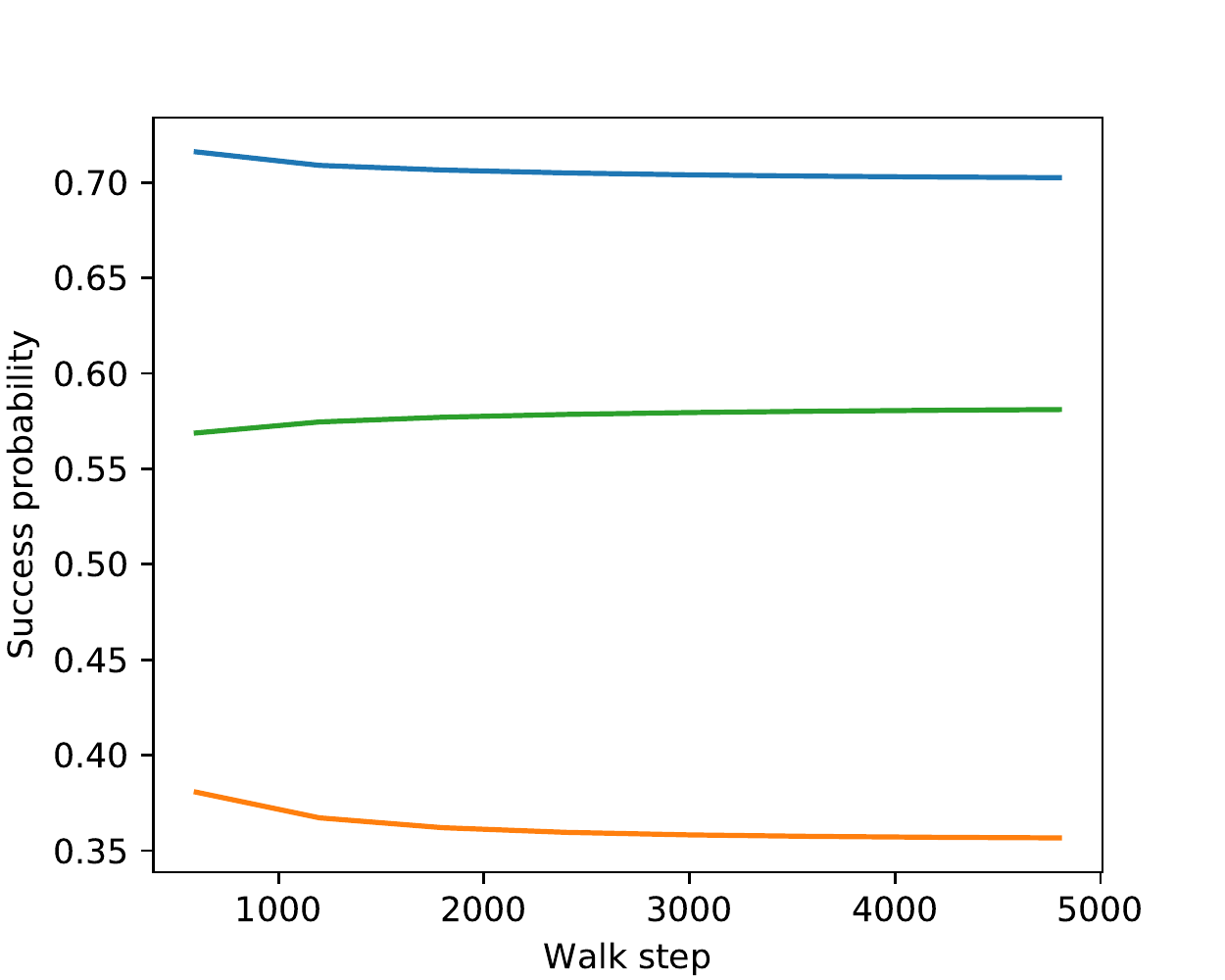}}
\caption{Distributed information scheme classification success  probability evolution in time in
	scenarios 1 (a) and 2 (b) for class A (blue), B (red) and C
	(green)}\label{fig:results-walk}
\end{figure}

In Fig.~\ref{fig:results-walk} the results obtained for the introduced scenarios
are presented. In the case when the training samples are mixed, the
probabilities for the very start are close to the limiting ones, but it still
takes hundreds of steps to reach the exact numbers with high precision.
In the case when the samples are grouped into clusters from the same class the
results are highly disturbed in the beginning. The class that we start from is
chosen relatively often, what makes success rate for that class higher than
normal, but the results for other classes are significantly worse.
Let us note that for given cycle size any node is obtainable in 75 steps.
Thus, one could consider other walk dynamics, in particular a coherent quantum
walk, that features ballistic spreading, for a more efficient protocol.

\subsection{Classification with quantum state recycling}

The model introduced so far harnessed the information in all of the nodes in
limited sense. The walk was designed to make it possible to measure any of the
nodes, thus make a comparison to any of the training samples. This allows to
obtain the same classification probability as in the initial proposal, but does
not harness the information encoded in the state after the measurement. So
called state recycling proved useful in quantum
algorithms~\cite{martin2012experimental} and with a
model of sequences of generalized measurements we can study possibilities of
harnessing this information.

\begin{figure}[h!]
	\begin{center}
		\includegraphics[scale=1.]{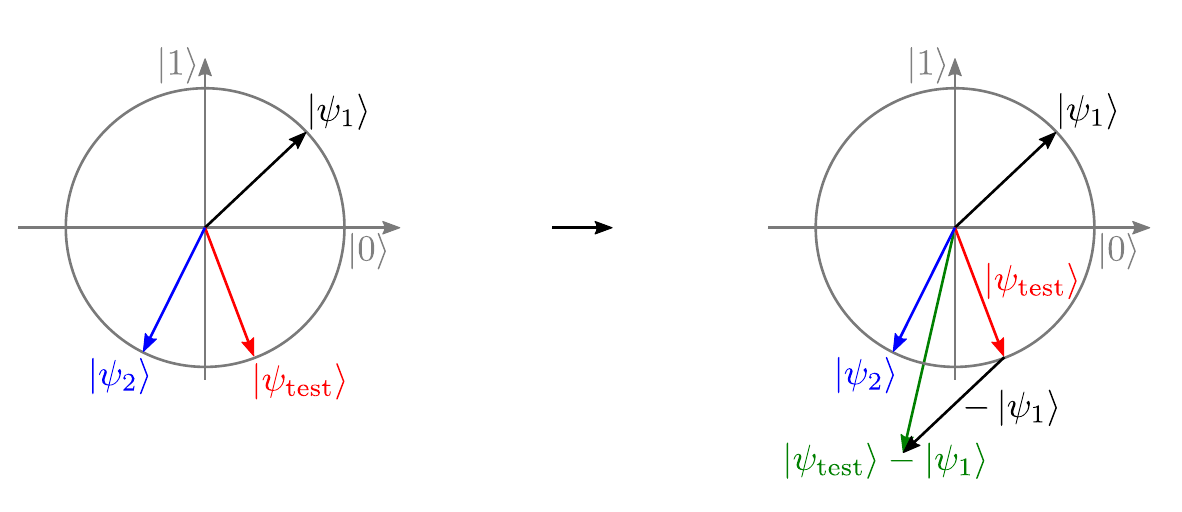}
	\end{center}
\caption{Classification measurement impact on the feature register for a test
state $\ket{\psi_{\mathrm{test}}}$ compared to reference state $\ket{\psi_{1}}$
resulting in preparation of a $\ket{\psi_{\mathrm{test}}}-\ket{\psi_{1}}$
		state rotated towards ground truth class instance state $\ket{\psi_2}$}\label{fig:recycling}
\end{figure}

Note that if the measurement result indicates that the training element $m$
corresponding to the walk position during measurement does not belong to the
same class as the test element, the state after the measurement
($\test-\psim$) is rotated away from the state encoding element $m$ and
usually gets closer to the goal class. An example is presented in
Fig.~\ref{fig:recycling}.
It is
straightforward to check that rotation introduced by the measurement is
\begin{equation}
\phi \rightarrow	\phi' = \phi/2 + \pi/2, \Delta\phi = \pi/2-\phi/2,
\end{equation}
where $\phi$ is the (acute) angle between $\psim$ and $\test$. Thus the
resulting state is rotated towards the other pole by at least $\pi/2$ and
proportionally to the initial angle.

This behavior can be used to improve the classification algorithm. The whole
procedure will now consist of a sequence of steps, with measurements at each
step. The measurement indicated whether current position is a good fit for the
test element. If it is, the procedure stops. If not, it continues preserving the
information encoded in the state as follows.
\begin{enumerate}
	\item For given test sample compute normalized vector $x_{\mathrm{test}}$.
	\item Pick random training sample $i$ from a uniform distribution.
	\item Prepare a circuit that prepares the comparison state $(H\otimes1)\ket{\psi(i)}$.
	\item Run the QPU, preparing $(H\otimes1)\ket{\psi(i)}$ and measure the resulting state.
	\item If the result at the auxiliary register is 0, continue.
	 While the result at the auxiliary register is 1, repeat:
	\begin{itemize}
		\item Preserve the resulting state $\ket{\psi_{\mathrm{rotated}}}$.
		\item Pick random sample $j$ from a uniform distribution.
		\item Conditionally prepare sample state $\ket{\psi_j}$ resulting in\\
		$\ket{\psi}=\frac{1}{2}(\ket{0}\otimes\ket{\psi_{\mathrm{rotated}}}+\ket{1}\otimes\ket{\psi_j})$.
		\item Measure the state $(H\otimes1)\ket{\psi}$.
	\end{itemize}
	\item Return $y_j$.
\end{enumerate}

The efficiency can be additionally affected with walk graph manipulation. When
the labels of the element in the training set are known, we propose to use a
graph where edges exist only between nodes belonging to different classes. In a
simple 2 class case this results in a bipartite graph.

\subsection{Quantum state recycling example}
We analyze the efficiency of the proposed  scheme with an experiment, in which
we compare its successful classification probability with the basic scheme. For
simplification we take only 2 classes of vectors from the same dataset as
before. We consider a set of transition operators, one for each pair of the
training elements defining an OQW. Then, we consider two scenarios:
\begin{itemize}
\item one classification measurement with random training sample,
\item a sequence of up to two steps, with the second one done only
if the result of the measurement during the first one is negative. The second
measurement is acting on the state resulting from the first one.
\end{itemize}
In this case the underlying OQW is done on a complete graph, stopping after
first successful comparison. 

\begin{table}[h]
	\begin{center}
		\begin{tabular}{|c|c|c|}
			\hline
			\multicolumn{3}{|c|}{4 features} \\ \hline
			Class & 1 step & 2 steps  \\ \hline
			A & 0.722 & 0.734  \\ \hline
			B & 0.689 & 0.710 \\ \hline
		\end{tabular}
	\end{center}
	\caption{Average successful classification probability with and without state recycling resulting from leave one out cross-validation on 2 classes chosen from the Iris dataset}\label{fig:recycling-results}
\end{table}

We measure the classification efficiency with leave one out cross validation.
The results are presented in Table~\ref{fig:recycling-results}. For the
considered data  the state recycling scheme provides better overall average
success probability. Moreover the success probability was better for $94\%$ and
$86\%$ of the samples for class A and B respectively.

\section{Discussion}

The results presented in this work open new possibilities to develop 
distance-based classification methods for near-term quantum processing units.

The most basic consequence of the introduced hybrid classical-quantum method is
parallel computation potential.
Given that only a part of the training information undergoes coherent evolution,
a number of small QPUs can be used to obtain higher classification accuracy.

Moreover, joining quantum cryptographic protocols with distributed information
classification could bring novel applications for QPUs. Harnessing
information routing based on quantum walks can cause non-trivial dynamics.
The walks introduced in this paper converge to the uniform distribution, but in
general quantum walks can feature non-trivial probability
distributions~\cite{konno2008quantum}. In particular, it has been shown that
additional connections, while reducing distances, can cause periodic in space,
non-uniform limiting behavior, in result leading to disturbed classification
probabilities~\cite{sadowski2016lively}. Similarly, non-random controlled walk
direction evolution leads to phenomena that is not present in classical systems,
resulting in game-like behavior that requires certain strategy to assure
that all of the points are properly included~\cite{miszczak2014magnus}.

Finally, we have shown that quantum implementation of a distance-based
classifier can achieve better accuracy when a state after the final measurement
is preserved and processed. This leads to a number of open questions.
The foundations of the observed improvement will be an object of our further
study.
Quantum processing units are naturally well suited for processing
high-dimensional vectors in large tensor product spaces, which should boost
their time performance.
Thus, significant additional performance improvement in terms of accuracy,
especially with relatively small
near-terms devices, would be a very desirable result.

\section{Acknowledgements}
Financial support by the Polish National Science Centre under project number
2015/17/B/ST6/01872 is gratefully acknowledged.

\bibliographystyle{unsrt}
\bibliography{ml}

\end{document}